\documentclass[english,12pt,aps,prd,a4paper,preprintnumbers,floatfix,nofootinbib,showpacs,superscriptaddress, notitlepage]{revtex4-1}
\pdfoutput=1
\usepackage[usenames,dvipsnames]{color}  
\usepackage{graphicx}
\usepackage{caption}
\usepackage{subcaption}
\usepackage{amsmath}
\usepackage{amssymb}
\usepackage{enumitem}
\usepackage[colorlinks=true,citecolor=darkgreen,urlcolor=darkgreen, pdfborder={0 0 0}]{hyperref}
\usepackage[normalem]{ulem}
\usepackage{hyperref}
\hypersetup{colorlinks=true,citecolor=green,linkcolor=blue,urlcolor=blue}
\usepackage{soul}

\def \Nucl{\mathcal N}
\newcommand{\bwt}{\begin{widetext}}
\newcommand{\ewt}{\end{widetext}}
\newcommand{\newc}{\newcommand}
\newc{\hc}{\dagger}
\newc{\pd}{\partial}
\newc{\beq}{\begin{equation}}
\newc{\eeq}{\end{equation}}
\newc{\beqa}{\begin{eqnarray}}
\newc{\eeqa}{\end{eqnarray}}
\newc{\bi}{\begin{itemize}}
\newc{\ei}{\end{itemize}}
\newc{\ra}{\rightarrow}
\newc{\la}{\leftarrow}
\newc{\lra}{\longrightarrow}
\newc{\lla}{\longleftarrow}
\newc{\Lra}{\Longrightarrow}
\newc{\Lla}{\Longleftarrow}
\newc{\half}{\frac{1}{2}}
\newc{\fth}{\frac{1}{4}}
\newc{\del}{\delta}
\newc{\Del}{\Delta}
\newc{\gm}{\gamma}
\newc{\Gm}{\Gamma}
\newc{\lam}{\lambda}
\newc{\kap}{\kappa}
\newc{\tri}{\triangle}
\newc{\eps}{\epsilon}
\newc{\epsp}{\epsilon^\prime}
\newc{\wt}{\widetilde}
\newc{\ovl}{\overline}
\newc{\tchi}{\tilde{\chi}}
\newc{\ds}{\displaystyle}
\newc{\pmt}{\pm\!\pm}
\newc{\PL}{\hat{L}}
\newc{\PR}{\hat{R}}

\newc{\msm}{\mathrm{SM}}
\newc{\msh}{\mathrm{sh}}
\newc{\mtev}{\mathrm{TeV}}
\newc{\mgev}{\mathrm{GeV}}
\newc{\mmev}{\mathrm{MeV}}
\newc{\mkev}{\mathrm{keV}}
\newc{\mev}{\mathrm{eV}}
\newc{\Tr}{\mathrm{Tr}}
\newc{\nonr}{\nonumber}
\newc{\clbl}{\color{blue}}
\newc{\clg}{\color{green}}
\newc{\clr}{\color{red}}
\mathchardef\mhyphen="2D
\newc{\SL}{\not\!\!}

\begin{document}
\title{ Constraints on light singlet fermion interactions from coherent elastic neutrino-nucleus scattering}
\date{\today}

\author{We-Fu Chang}
\email{wfchang@phys.nthu.edu.tw}
\affiliation{Department of Physics, National Tsing Hua University, No. 101, Section 2, Kuang-Fu Road, Hsinchu, Taiwan 30013, R.O.C.}
\author{Jiajun Liao}
\email{liaojiajun@mail.sysu.edu.cn}
\affiliation{School of Physics, Sun Yat-Sen University, Guangzhou, 510275, China}
\begin{abstract}
 The exotic singlet fermions $\chi$, with a mass $m_\chi\lesssim 50$ MeV, could be produced at the  coherent elastic neutrino-nucleus scattering (CE$\nu$NS) experiments through the
$\nu  \Nucl \rightarrow \chi \Nucl$ process. Due to the coherent enhancement, it offers a unique way to study how $\chi$ interacts with the Standard Model (SM) sector.
Based on the most general dimension-6 effective Lagrangian, we perform a comprehensive study on the relevant interaction between $\chi$ and the SM sector.
From the current and future COHERENT and future CONUS experiments, we obtain the upper bounds on the Wilson coefficients  for the dipole, scalar, vector, and tensor interactions.  For $m_\chi $ below 10~MeV, future CONUS data has the best sensitivity, while for $m_\chi$ between 10 MeV$-50$~MeV, the current and future COHERENT bounds dominate.
These limits are complementary to those from neutrino oscillation and collider searches. Moreover, the bounds do not depend on the charge conjugation property of $\chi$, nor whether $\chi$ is dark matter or not.

\end{abstract}
\maketitle

\newpage
\section{Introduction}

Singlet fermions, collectively denoted as $\chi$ in this paper, are gauge singlets under the  Standard Model (SM) $SU(3)_c\times SU(2)_L\times U(1)_Y$  symmetries. $\chi$ is widely discussed in models of new physics beyond the SM. To name a few, the singlet fermion(s) could play the role as the sterile neutrino(s) in the neutrino mass generation\cite{Minkowski:1977sc, Glashow:1979nm,GellMann:1980vs, Yanagida:1980xy, Mohapatra:1979ia, Schechter:1980gr, Mohapatra:1986aw, Mohapatra:1986bd}.
$\chi$ is also a popular candidate to account for the anomalies observed at short baseline neutrino oscillations\cite{AguilarArevalo:2010wv, Aguilar-Arevalo:2012fmn, Aguilar:2001ty}.
Moreover, if the lightest one of $\chi$ is stable or cosmologically long-lived, it could be a viable dark matter (DM) candidate\cite{Adhikari:2016bei, Boyarsky:2018tvu}.   The physical mass of $\chi$ is highly model dependent, and  allows to vary in this paper. Motivated by the possible sterile neutrino warm DM and many neutrino experiments at low energies, in this work, we only consider the cases that $\chi$ is much lighter than the SM electroweak scale.

In the ultra-violet (UV) theory, if $\chi$ does not carry any  beyond SM quantum number, the renormalizable dim-4 Yukawa coupling term $\bar{L} H \chi$ is allowed by the SM symmetries, where $L$ and $H$ are the SM lepton and Higgs doublet, respectively.   Below the SM electroweak spontaneous symmetry breaking (SSB) scale, the SM Higgs acquires a nonzero vacuum expectation value(VEV), $v_H\sim 246$~GeV, and bestows the Dirac mass connecting the SM neutrino and $\chi$. The Dirac mass term leads to the mixings between $\chi$ and the SM neutrinos, and it is crucial for both the Dirac and the seesaw-type neutrino mass generation mechanisms.
 Also, $\chi$  can participate in the SM neutral/charged current (NC/CC) interactions through
the mixing with the SM neutrinos.
In addition to the $\bar{L} H \chi$ Yukawa term, it is also possible to have new scalar couplings through the Higgs portal, where the SM Higgs mixes with the potentially light exotic scalar field(s).
Usually, the scalar couplings of $\chi$ to the SM Higgs or the exotic scalar boson(s) are suppressed by the active neutrino mass, and thus negligible.

On the other hand, if $\chi$ is charged under some symmetry $G_{BSM}$ beyond the SM, the above Yukawa term is forbidden in the UV theory.  But the effective couplings with the SM neutrinos can be achieved if the symmetry $G_{BSM}$ is broken and the proper mediator exists.  In this case, the mediator(s) could yield new interactions other than the SM NC/CC interactions as well.
Moreover, the additional interactions  are not necessarily negligible comparing to the dominate two, $\chi\mhyphen\nu\mhyphen Z^0$ and $\chi\mhyphen l^-\mhyphen W^+\,(l=e,\mu,\tau)$, through the $\chi\mhyphen\nu$ mixings.
As an example, the new $\chi\mhyphen\nu$ interactions emerge in a 3-portal model recently discussed in\cite{Chang:2019sel}.
This simple model consists of a pair of vector fermions, $\chi_{L/R}$, and a singlet scalar, $\phi_x$. They are both charged under a hidden gauged $U(1)_x$.
It also employs the sterile neutrinos, $N_R$'s, for the type-1 see-saw neutrino mass generation. The $N_R$'s are assumed to be heavy, around the typical lepton number violating scale $M_L \sim {\cal O}(10^{12-14})$ GeV, as in the standard high scale type-1 see-saw models. The $U(1)_x$ is SSB when $\phi_x$ acquires a VEV, $v_x$, and $M_L\gg v_x \gg v_H$ is assumed.
 In addition to the Weinberg operator $(LH)^2$, more effective operators emerge simultaneously after integrating out $N_R$'s.
For instance, the new $(LH)(\ovl{\chi^c}_L \phi_x^*)$ operator couples the singlet fermion to the SM leptons. Moreover, it leads to a dim-4 $\ovl{\chi^c}_L LH$ effective interaction by replacing $\phi_x$ by its VEV $v_x$. Since $v_x\gg v_H$, the scalar coupling in this model is much larger than the traditional one, which stems from the $\chi\mhyphen\nu$ mixing alone.
In addition to the neutrino portal $\chi\mhyphen\nu\mhyphen Z^0$ NC interaction, there are more interactions from the Higgs-portal and the $U(1)_Y\mhyphen U(1)_x$ kinematical mixing gauge portal as well.

For a general discussion below the electroweak scale, we use $\theta_{\chi i}$  to denote the unknown mixing angle, regardless of its UV origin, between $\chi$ and the $i$-flavor SM neutrino.
Depending on its mass $m_\chi$ and $\theta_{\chi i}$, the singlet fermion can be probed at the neutrino oscillation experiments, the spectrum endpoint in the beta decays, colliders, or the lepton universality tests, and so on\cite{Dragoun:2015oja, deGouvea:2015euy, Atre:2009rg, Bryman:2019ssi, Deppisch:2015qwa, Bolton:2019pcu}.
The coherent elastic neutrino-nucleus scattering (CE$\nu$NS), $\nu \Nucl \ra \nu \Nucl$, predicted by the SM\cite{Freedman:1973yd} has been observed by the COHERENT collaboration in a cesium iodide (CsI) detector~\cite{Akimov:2017ade}, and recently confirmed in a liquid argon (LAr) detector~\cite{Akimov:2020pdx}.
Other experiments, which include CONUS~\cite{CONUS_exp, Buck:2020opf}, $\nu$-cleus~\cite{Strauss:2017cuu}, CONNIE~\cite{Aguilar-Arevalo:2016khx}, MINER~\cite{Agnolet:2016zir}, TEXONO~\cite{Wong:2010zzc}, $\nu$GEN~\cite{Belov:2015ufh}, and Ricochet~\cite{Billard:2016giu}, also plan to measure CE$\nu$NS in the near future.

 The measurement of CE$\nu$NS opens up a new avenue to explore the new physics associated with $\chi$.
Given a nonzero mixing between $\chi$ and the SM neutrino, the relevant effective low energy 4-fermi operators, $(\bar{\nu}\gamma^\mu_L \chi)(\bar{q}\gamma_\mu q)+h.c.$, can be generated by the SM NC interaction.
If the energy transfer is much smaller than the nucleus size inverse, the elastic scattering cross section will be coherently enhanced. Then any light enough singlet fermion(s), not limited to the one which serves as the dark matter, could be produced in the final state by the process $\nu \Nucl \ra \chi\Nucl$ or $\bar{\nu} \Nucl \ra \chi^c \Nucl$ within the same experiment setup designed to study the SM CE$\nu$NS process\footnote{
The inverse process $ \chi \Nucl \ra \nu \Nucl$ was recently discussed for the novel detection of the fermionic DM\cite{Dror:2019onn,Dror:2019dib}, and could be potentially constrained by the CE$\nu$NS experiments. The process $ \chi \Nucl \ra \chi \Nucl$ with $\chi$ serving as a dark matter candidate has been studied in the CE$\nu$NS experiments~\cite{deNiverville:2015mwa, Ge:2017mcq, Dutta:2019nbn}.}.
As mentioned earlier, the UV theory could potentially generate more effective operators with Lorentz structures different from the NC one just discussed. To account for this possibility, we consider the most general model-independent set of  4-fermi operators.
 In this work, we should study how the minimal set of 4-fermi operators impacts CE$\nu$NS with $\chi$ in the final state.

The most general model-independent dim-6 effective Lagrangian will be considered in Sec.\ref{sec:EFT},  followed by the discussion of tree-level CE$\nu$NS cross-section and the nucleus form factors in Sec.\ref{sec:XS}. Section \ref{sec:exp_bound} is devoted to the current and future constraints on the Wilson coefficients derived from the current and future COHERENT and CONUS experiments.
We summarize our results in Sec.\ref{sec:discuss}. Some calculation details are collected in Appendix~\ref{apd:amplitude}, and a UV complete model is presented in Appendix~\ref{apd:UVmodel}.

\section{Effective operators}
\label{sec:EFT}
Since the momentum transfer squared, $-q^2$, involved in the neutrino-nucleus $\nu \Nucl \ra \chi \Nucl$ coherent scattering  are relatively small, i.e., $-q^2\ll (\text{GeV})^2$, it makes sense for one to consider the effective theory below the electroweak scale. In this energy scale, all the degrees of freedom heavier than electroweak scale have been integrated out; even the SM $Z,W^\pm$ bosons and top quark are absent in the effective theory. Therefore, it is natural to set the cutoff at the electroweak scale.
The most general $SU(3)_{QCD}\times U(1)_{QED}$ invariant dim-6 effective Lagrangian\footnote{ Above the electroweak symmetry breaking, the gauge invariant dipole interaction is dim-6, $ \bar{L}\sigma^{\mu\nu}(a_M +ia_E \gamma^5)\chi H  B_{\mu\nu}$,  where $B_{\mu\nu}$ is the field strength of $U(1)_Y$, $H$ and $L$ are the SM Higgs and lepton doublets, respectively.
Below the electroweak scale, this term contributes to both the dipole and tensor interactions. For $m_\chi<m_{\pi^0}$,  $\chi\ra \nu \gamma$ is the only 2-body decay mode of $\chi$ decaying into SM particles, allowed by the Lorentz and the $SU(3)_c\times U(1)_{QED}$ symmetries, independent of any model.   } can be parametrized as
\beqa
\frac{\sqrt{2}}{G_F} {\cal L}  &=& \frac{v_H}{\sqrt{2}}\left[\bar{\nu}\sigma^{\mu\nu}(a_M +ia_E \gamma^5)\chi\right]F_{\mu\nu}
+\left[\bar{\nu}(C_S^q+i\gamma^5 D_P^q)\chi\right][\bar{q}q]\nonr\\
&&
+\left[\bar{\nu}\gamma^\mu(C_V^q+\gamma^5 D_A^q)\chi\right][\bar{q}\gamma_\mu q]
+\left[\bar{\nu}\sigma^{\mu\nu}C_T^q \chi\right][\bar{q}\sigma_{\mu\nu}q] + \text{h.c.}\,,
\label{eq:H_eff}
\eeqa
where $v_H\simeq 246$ GeV is the vacuum expectation value of the SM Higgs.
Note that we use the above convention such that in the Hermitian conjugation all the dimensionless coefficients take their complex conjugations but the signs in the Lagrangian remain unchanged.
 Here we have dropped all $\gamma^5$-terms associated with the quark, which do not receive coherent enhancement in the low energy $\nu \Nucl \ra \chi\Nucl$ elastic scattering. Due to the identity that $\sigma^{ab}\gamma_5= \frac{i}{2}\epsilon^{abcd}\sigma_{cd}$, the $\gamma^5$ associated with leptons in the tensor term, $\left[\bar{\nu}\sigma^{\mu\nu}\gamma^5\chi\right][\bar{q}\sigma_{\mu\nu}q]$ can be shifted to the quark side. Therefore, this term is also  suppressed by the average nucleon spin, and thus ignored.
For anti-neutrino, the coefficients $\{a_M,a_E, C_S,D_P,C_V, D_A, C_T \}$ should be replaced by
$\{-a_M^*,-a_E^*, C_S^*, D_P^*, -C_V^*, D_A^*, -C_T^* \}$ by performing charge conjugation to the Lagrangian.
 Moreover, for the scattering process with $\chi$ produced in the final state, without being detected, its charge conjugation properties do not got involved, so despite whether $\chi$ is Majorana or Dirac, our discussion applies to both cases.

In general, the unknown coefficients $C$'s and $D$'s could be quark and neutrino flavor dependent. Here the flavor indices are suppressed, and they will be specified only when needed.
Note that the contact interaction description is no longer valid for a light bosonic mediator\footnote{See \cite{Brdar:2018qqj} for a recent discussion.} of mass $m_X \lesssim 50$ MeV, the typical momentum transfer in $\nu-\Nucl$ coherent scattering experiment using a stopped pion decay source.
 In this paper, we are only interested in the cases that $m_X$ is much larger than the momentum transfer, and
our result can be easily translated to set bounds on the strength of coupling-mass ratio for $ m_X\gg 50$ MeV.

\section{scattering cross-section}
\label{sec:XS}
\begin{figure}[htb]
\centering
\includegraphics[width=0.6\textwidth]{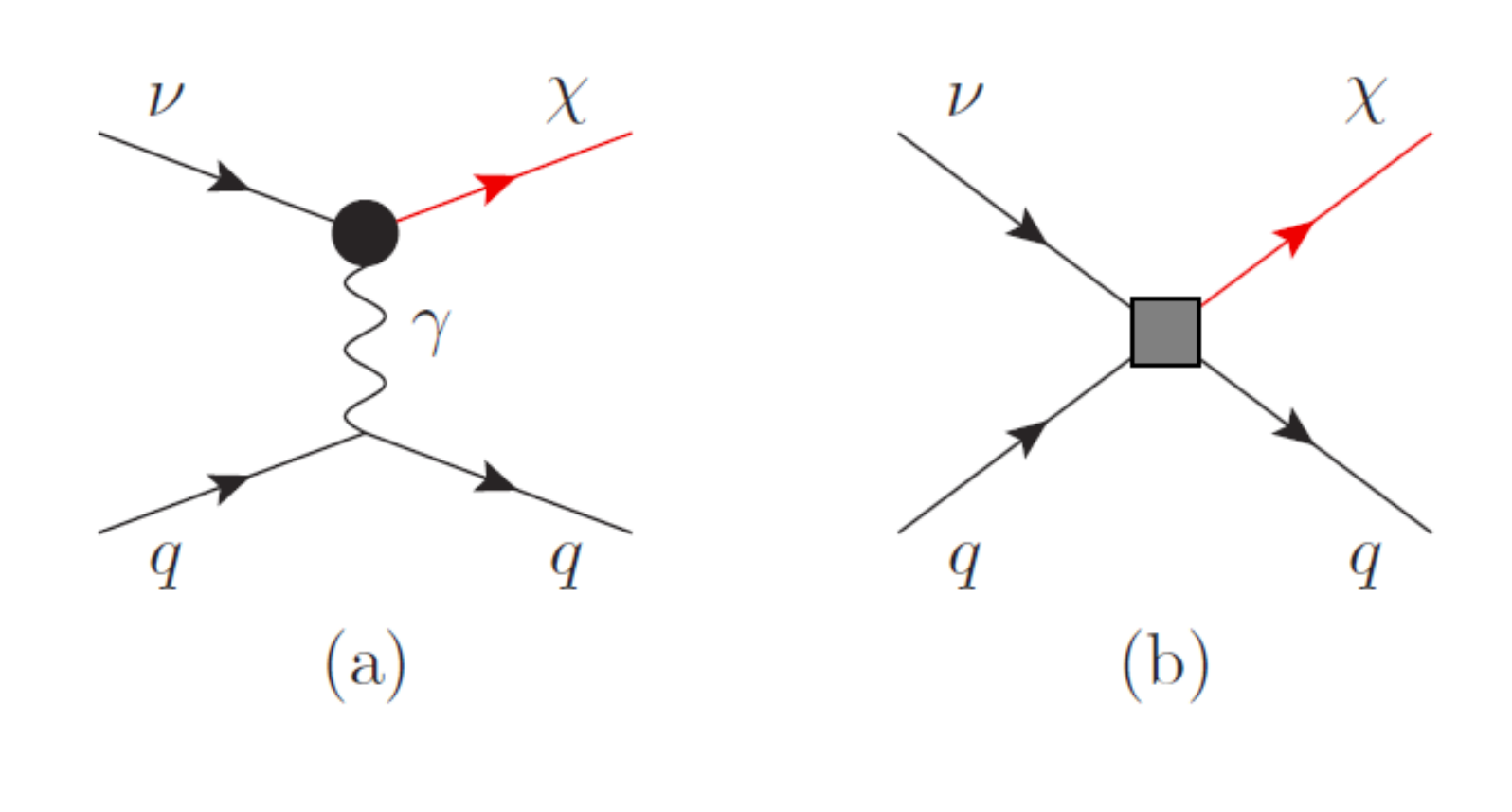}
\caption{The tree-level Feynman diagrams for the $\nu q \ra \chi q$ coherent scattering process, where the circular bulb and the square  represent the effective dipole interaction and the dim-6 4-Fermi interaction, respectively.}
\label{fig:TLFD}
\end{figure}
The effective Lagrangian, Eq.(\ref{eq:H_eff}), gives rise to the tree-level neutrino-quark  elastic scattering,
see Fig.\ref{fig:TLFD}. The corresponding  differential cross-section can be easily calculated\footnote{See Appendix~\ref{apd:amplitude} for the scattering matrix elements.}
and converted into the one for neutrino-nucleus coherent elastic scattering.
Taking into account of both the hadronic form factors for quarks in nucleons and the nuclear form factors for nucleons in nuclei, the differential cross section of $\nu  \Nucl \rightarrow  \chi \Nucl $ can be written as
\beqa
\frac{d \sigma}{d T} &=& \frac{G_F^2 M }{4\pi} (|C_S(q^2)|^2+|D_P(q^2)|^2)\left(1+\frac{T}{2M}\right)\left( \frac{M T}{E^2}+ \frac{m_\chi^2}{2E^2} \right)
\nonr\\
&+& \frac{G_F^2 M }{2\pi} (|C_V(q^2)|^2+|D_A(q^2)|^2)\left[ 1-\frac{T}{E}+\frac{T}{2E^2}(T-M) -\frac{m_\chi^2}{4E^2} \left(1+\frac{2E}{M}-\frac{T}{M}\right)\right]
\nonr\\
&+& \frac{4 G_F^2 M }{\pi} |C_T(q^2)|^2 \left[ 1-\frac{T}{E}+\frac{T}{4E^2}(T-M) -\frac{m_\chi^2}{4E^2} \left(\frac{1}{2}+\frac{2E}{M}-\frac{T}{2M}\right)\right]\nonr\\
&+& \frac{ 2 G_F^2 Z^2 s_W^2 }{\pi}\left(\frac{M_W^2}{M}\right)  (|A_M(q^2)|^2+|A_E(q^2)|^2)\nonr\\
&&\times \left[ -\frac{M}{E}+ \frac{m_\chi^2}{4E^2} +\frac{M}{T}\left( 1-\frac{m_\chi^2}{4E^2} -\frac{m_\chi^2}{2 M E}+ \frac{m_\chi^4}{8 M^2 E^2}\right) -\frac{m_\chi^4}{8 T^2 E^2} \right]\nonr\\
&+& \frac{ G_F^2 Z s_W   }{\sqrt{2}\pi} \left(\frac{M M_W}{E}\right) \Re[C_T^*(q^2) A_M(q^2) ] \left[ \frac{2T}{E} - \frac{m_\chi^2}{M E} -\frac{m_\chi^4}{ T M^2 E} \right]\nonr\\
&-&  \frac{ \sqrt{2} G_F^2 Z s_W   }{\pi} \left(\frac{M M_W}{E}\right)\Re[ (A_M^*(q^2) C_S(q^2)+ A_E^*(q^2) D_P(q^2))] \left[ 1 -\frac{T}{2E}- \frac{m_\chi^2}{4 M E}  \right]\nonr\\
&-& \frac{ G_F^2 M }{2\pi} \Re[C_T^*(q^2) C_S(q^2)] \left[ \frac{2T}{E} -\frac{T^2}{E^2}- \frac{m_\chi^2 T}{2 M E^2}  \right]\,,
\label{eq:xsec}
\eeqa
where $s_W$ is the short hand for the weak mixing, $\sin\theta_W$, $E$ the energy of incoming neutrino, $T$ the recoil energy of nucleus, and $M$ the mass of the target nucleus.
The $q^2$-dependent effective neutrino-nucleus couplings are related to the fundamental neutrino-quark couplings as follows~\cite{AristizabalSierra:2018eqm, AristizabalSierra:2019zmy}:
\begin{align}
\label{eq:scalar-CS-FF}
C_S(q^2)&=\sum_{q=u,d}C_s^{q}\left[N\frac{m_n}{m_q}f_{T_q}^nF_n(q^2)
+ Z\frac{m_p}{m_q}f_{T_q}^pF_p(q^2)\right]\ ,
\\
C_V(q^2)&=N(C_V^u+2C_V^d)F_n(q^2) + Z(2C_V^u+C_V^d)F_p(q^2)\ ,
\\
C_T(q^2)&=N(\delta_u^nC_T^u+\delta_d^nC_T^d)F_n(q^2)
+ Z(\delta_u^pC_T^u+\delta_d^pC_T^d)F_p(q^2)\ ,
\\
A_M(q^2)&= a_MF_p(q^2)\,,
\end{align}
 where $Z$ ($N$) is the number of protons (neutrons) in the nucleus,  $f_{T_q}^{p}$ ($f_{T_q}^{n}$) the fraction of nucleon mass contributed by a given quark flavor $q$, $\delta_q^{p}$ ($\delta_q^{n}$) the tensor charges, and $F_p(q^2)$ ($F_n(q^2)$) the nuclear form factor for protons (neutrons). Here we adopt the Helm form factors~\cite{Helm:1956zz} in our analysis, and assume the neutron form factor is the same as proton. Note that deviations from this assumption are possible due to uncertainties on the root-mean-square radius of the neutron distribution~\cite{AristizabalSierra:2019zmy}. For the scalar and  tensor parameters, we use the following values~\cite{AristizabalSierra:2018eqm}:
\begin{alignat}{4}
\label{eq:FF-values}
f_{T_u}^p&=0.019\ ,\qquad&f_{T_d}^p=0.041\ , \qquad
f_{T_u}^n&=0.023\ ,\qquad&f_{T_d}^n=0.034\ ,
\nonumber\\
\delta_u^p&=0.54\ ,\qquad&\delta_d^p=-0.23\ , \qquad
\delta_u^n&=-0.23\ ,\qquad&\delta_d^n=0.54\ ,
\end{alignat}
which are taken from Refs.~\cite{Jungman:1995df, Anselmino:2008jk}.
The expression for $D_P$ ($D_A$) [$A_E$] can be obtained by trading $C_S^{q}\rightarrow D_P^{q}$ ($C_V^{q}\rightarrow D_A^{q}$) [$a_M\rightarrow a_E$].
Note that each photon propagator from the dipole term gives one $1/T$ proportionality. Since we are not dealing with the UV model, this IR divergence is expected. However, we are not concerned about this IR divergence because the experiments are not sensitive to such low $T$ regions.

Since the final states are different from the initial states, there is no interference between the  $\nu \Nucl\ra \chi \Nucl$ and the SM $\nu \Nucl\ra \nu \Nucl$ processes. Because neither SM $\nu$ nor singlet $\chi$ is detected in the scattering,  the total coherent scattering cross-section is the sum of Eq.~(\ref{eq:xsec}) and the SM one.  Moreover, due to the same chirality of final states, the dipole, scalar, and tensor interactions can mix and yield different interference patterns. The last two interference terms in  Eq.~(\ref{eq:xsec}) change signs when the incoming neutrino is replaced by anti-neutrino.
As shown in Ref.~\cite{AristizabalSierra:2019ufd}, for a light vector mediator scenario, due to the interference between the new vector and the SM interactions, the presence of a dip in the recoil spectrum in some parameter regions can be used to constrain the CP violating effects in future CE$\nu$NS experiments. In our scenarios, since there is no interference between the SM and new interactions, there is no dip in the event rate spectrum as compared to the SM predictions. From Eq.~(\ref{eq:xsec}), we see that the differential cross section is also affected by the nonzero CP-violating phases of the dipole, scalar, and tensor Wilson coefficients due to the interference terms between them. These nonzero CP-violating  phases can in principle be probed at future CE$\nu$NS experiments if a high precision measurement of the  event rate spectrum is obtained. However, since the interference terms depend on two new interactions, it is more difficult to distinguish the CP-violating effects from the CP-conserving case unless both of the new interactions are large.
The Lorentz structures of the interactions not only affect the matrix elements of the cross section, but also modify the corresponding form factors.  Therefore, in principle, each of the dipole, scalar, and tensor coefficients can be disentangled with precisely measured differential cross-sections on various targets from both the neutrino and anti-neutrino sources in the future.

\begin{figure}[t]
	\centering
	\includegraphics[width=0.6\textwidth]{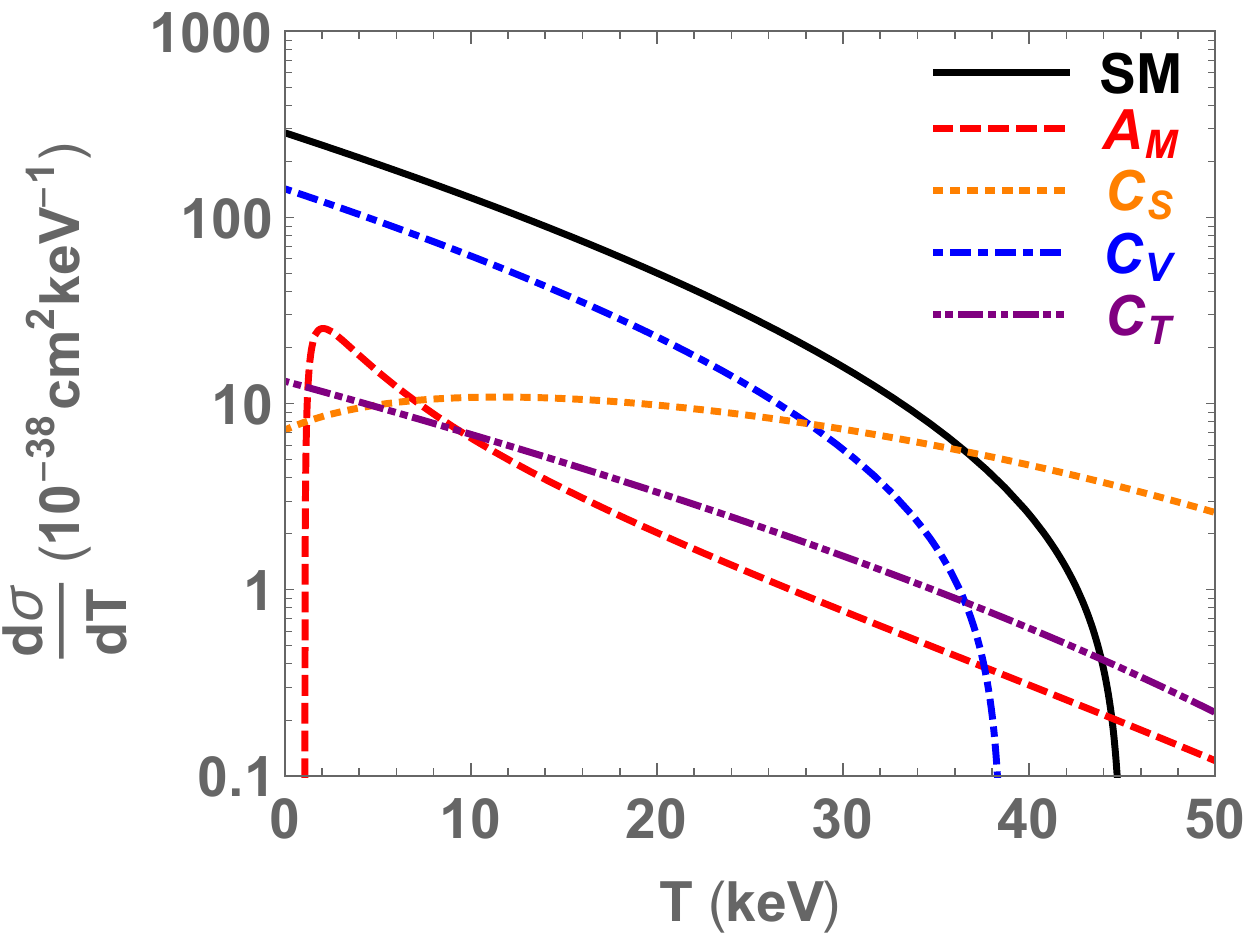}
	\caption{ The differential cross sections of $\nu \Nucl\ra \chi \Nucl$ as a function of the nuclear recoil energy.  The red dashed (orange dotted) [blue dot-dashed] \{purple dot-dot-dashed\} curve corresponds to the dipole (scalar) [vector] \{tensor\} interaction with $a_M=10^{-4}$ ($C_S^u=C_S^d=10^{-2}$) [$C_V^u=C_V^d=0.1$] \{$C_T^u=C_T^d=0.1$\}. Here we take the target nucleus as $^{133}$Cs, and assume $m_\chi=40$~MeV, $E_\nu=53$~MeV. The SM differential cross section of $\nu \Nucl\ra \nu \Nucl$ is also shown as the black solid curve for comparison. }
	\label{fig:xsection}
\end{figure}

We show a plot of the differential cross sections for different types of interactions as a function of the nuclear recoil energy for illustration. Here we take $^{133}$Cs as the target nucleus, and assume $m_\chi=40$~MeV, $E_\nu=m_\mu/2\approx53$~MeV. For the Wilson coefficient of each interaction, we consider four cases: (i) $a_M=10^{-4}$, (ii) $C_S^u=C_S^d=10^{-2}$, (iii) $C_V^u=C_V^d=0.1$, and (iv) $C_T^u=C_T^d=0.1$. The coefficients unmentioned in each case are assumed to be zero.   The differential cross sections of $\nu \Nucl\ra \chi \Nucl$ as a function of the nuclear recoil energy for the four cases are shown in Fig.~\ref{fig:xsection}.
The SM differential cross section of $\nu \Nucl\ra \nu \Nucl$ is also shown as the black solid curve for comparison.
 From Fig.~\ref{fig:xsection}, we see that the shape of the differential cross sections largely varies according to the types of interactions. For the dipole interaction, there is a peak at
 \begin{align}
 	  T_\text{peak}=\frac{m_\chi^4}{4ME^2}\left( 1-\frac{m_\chi^2}{4E^2} -\frac{m_\chi^2}{2 M E}+ \frac{m_\chi^4}{8 M^2 E^2}\right)^{-1}\sim 2\text{ keV}\,,
 \end{align}
 which can be derived from Eq.~(\ref{eq:xsec}).
Also, for the vector interaction, the overall shape of the differential cross section of $\nu \Nucl\ra \chi \Nucl$ is very similar to the SM case of $\nu \Nucl\ra \nu \Nucl$, which can be easily understood from Eq.~(\ref{eq:xsec}) since the two differential cross sections will only differ by an overall factor when  $m_\chi$ approaches 0.

Note that these new interactions will be further distinguished if the direction of the recoiling nucleus can be measured in future CE$\nu$NS experiments that use gaseous helium or fluorine as detector material~\cite{Abdullah:2020iiv}. The angular dependence of the differential cross section can be described as a $\delta$-function of $\cos\theta$~\cite{Abdullah:2020iiv, OHare:2015utx} with the expression of $\cos\theta$ given in Eq.~(\ref{eq:angle}).

\section{Constraints from CE$\nu$NS experiments}
\label{sec:exp_bound}
The COHERENT collaboration has observed CE$\nu$NS in a CsI detector at the 6.7$\sigma$ CL. The neutrinos measured at COHERENT are produced by the $\pi^+$ and $\mu^+$ decays at the Spallation Neutron Source (SNS) in the Oak Ridge National Laboratory~\cite{Akimov:2017ade}.  The energy distribution of the three neutrino flavors at SNS are well known and given by
\begin{align}
\label{eq:nu-spectra.COHERENT}
\phi_{\nu_\mu}(E_{\nu_\mu})&=\mathfrak{N}
\frac{2m_\pi}{m_\pi^2-m_\mu^2}\,
\delta\left(
1-\frac{2E_{\nu_\mu}m_\pi}{m_\pi^2-m_\mu^2}
\right) \ ,
\nonumber\\
\phi_{\nu_e}(E_{\nu_e})&=\mathfrak{N}
\frac{192}{m_\mu}
\left(\frac{E_{\nu_e}}{m_\mu}\right)^2
\left(\frac{1}{2}-\frac{E_{\nu_e}}{m_\mu}\right)\ ,
\nonumber\\
\phi_{\bar\nu_\mu}(E_{\bar\nu_\mu})&=\mathfrak{N}
\frac{64}{m_\mu}
\left(\frac{E_{\bar\nu_\mu}}{m_\mu}\right)^2
\left(\frac{3}{4}-\frac{E_{\bar\nu_\mu}}{m_\mu}\right)\,,
\end{align}
 where $ \mathfrak{N} =\frac{rtN_\text{POT}}{4\pi L^2}$ denotes the normalization factor with $r=0.08$ being the number of neutrinos per flavor produced per proton collision, $t$ the number of years of data collection, $N_\text{POT}=2.1\times 10^{23}$ the total number of protons delivered to the target per year, and $L$ the distance between the source and the detector~\cite{Akimov:2017ade}. Here $\nu_\mu$ is monochromatic with $E_{\nu_\mu}\approx30$~MeV, and the energies of $\nu_e$ and $\bar\nu_\mu$ are less than $m_\mu/2 \approx 53$~MeV.
The expected number of events with recoil energy in the energy range
 [$T$, $T+\Delta T$] can be calculated by
\begin{equation}
\label{eq:recoil-spectrum}
 N_{th} ( T )=\sum_{\alpha}\frac{m_\text{det}N_A}{M_{mol}}
\int_{\Delta T}\,dT \int_{E_\nu^\text{min}}^{E_\nu^\text{max}}\,dE_\nu\,\phi_\alpha(E_\nu)
\,\frac{d\sigma_{\alpha}}{d T}\, ,
\end{equation}
where $\alpha=\nu_\mu,\bar\nu_\mu, \nu_e$, $m_\text{det}$ is the detector mass, $M_{mol}$ the molar mass of the target nucleus,  and $N_A=6.022\times 10^{23}\,\text{mol}^{-1}$.
In the SM, the differential cross section for a given neutrino flavor $\nu_\alpha$ scattering off a nucleus is given by~\cite{Freedman:1973yd} ( also the SM limit of Eq.(\ref{eq:xsec}) ):
\begin{equation}
\label{eq:cevns-xsec}
\frac{d\sigma_\alpha}{d T}\approx\frac{G_F^2 M}{2\pi}
\left(2-\frac{T M}{E_\nu^2}\right)
\left[Ng_V^nF_N(q^2)+Zg_V^pF_Z(q^2)\right]^2\,,
\end{equation}
where $M$ is the mass of the target nucleus,  $g_p^V=\frac{1}{2}-2\sin^2\theta_W\approx 0.04$ and $g_n^V=-\frac{1}{2}$ are the SM weak couplings.

 To compare with the COHERENT data collected by the CsI detector, we convert the nuclear recoil energy to the photoelectrons (PEs) by using the relation $ n_\text{PE}=1.17( T /\text{keV}) $~\cite{Akimov:2017ade} \footnote{We do not use the new quenching factor given in Ref.~\cite{Collar:2019ihs} since it is still under investigation by the COHERENT collaboration~\cite{Konovalov}.}. We also utilize
the acceptance function given in Ref.~\cite{Akimov:2018vzs}, which is
\begin{equation}
\label{eq:acceptance}
\mathcal{A}(n_\text{PE})=\frac{k_1}{1+e^{-k_2(n_\text{PE}-x_0)}}\theta(n_\text{PE}-5)\,,
\end{equation}
where $k_1=0.6655$, $k_2=0.4942$, $x_0=10.8507$, $\theta(x)$ is the Heaviside function and
$n_\text{PE}$ the observed number of PEs.

For simplicity, we assume universal flavor-conserving couplings to quarks and neutrinos, and consider four cases with each of the four Wilson coefficients $\{a_M, C_S^q, C_V^q, C_T^q \}$ to be nonzero\footnote{From Eq.~(\ref{eq:xsec}), we see that the bound on $a_E$ ($D_P^q$) [$D_A^q$] is the same as that on  $a_M$ ($C_S^q$) [$C_V^q$].}.
Since the coefficients are flavor-independent, we do not use the timing information at the COHERENT experiment. To evaluate the statistical significance of a new interaction, we define
\begin{align}
\chi^2 = \sum_{i=4}^{15} \left[\frac{N_\text{meas}^i-N_\text{th}^i(1+\alpha)-B_\text{on}(1+\beta)}{\sigma_\text{stat}^i}\right]^2+\left(\frac{\alpha}{\sigma_\alpha}\right)^2+\left(\frac{\beta}{\sigma_\beta}\right)^2\,,
\end{align}
where $N_\text{meas}^i$ ($N_\text{th}^i$) is the number of measured (predicted) events per energy bin, $\alpha$ and $\beta$ are the nuisance parameters for the signal rate and the beam-on background with their uncertainties $\sigma_\alpha=0.28$ and $\sigma_\beta=0.25$~\cite{Akimov:2017ade}.  The statistical uncertainty per energy bin is determined by $\sigma_\text{stat}^i=\sqrt{ N_\text{meas}^i +2B_\text{SS}^i+B_\text{on}^i}$, where $B_\text{SS}$ is the estimated steady-state background from the anti-coincident (AC) data, and $B_\text{on}$ is the beam-on background from the prompt neutrons.

\begin{figure}[t]
	\centering
	\begin{subfigure}{.49\textwidth}
		\centering
		\includegraphics[width=\textwidth]{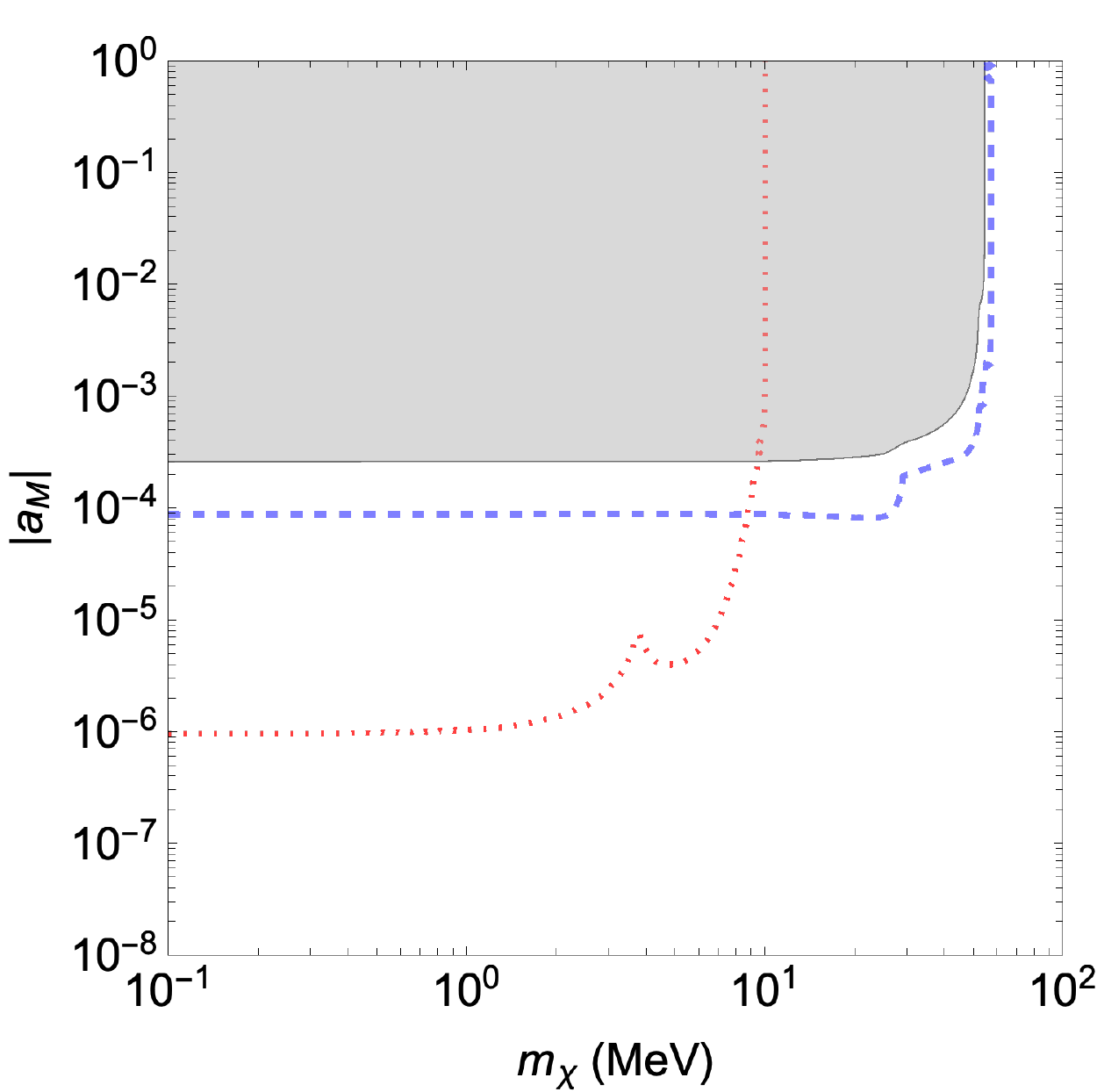}
		\label{fig:CS}
	\end{subfigure}
	\begin{subfigure}{.49\textwidth}
		\centering
		\includegraphics[width=\textwidth]{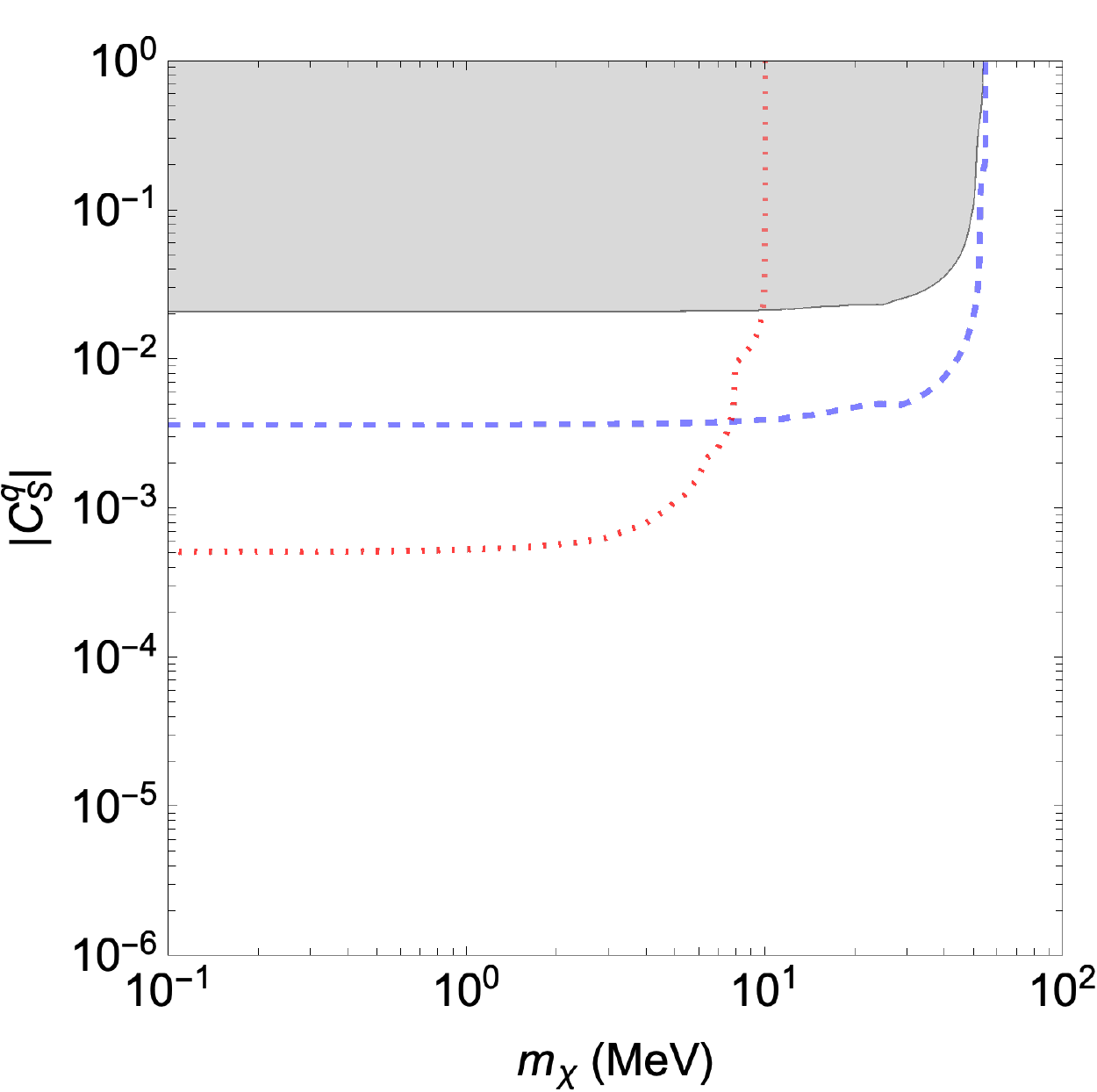}
		\label{fig:CV}
	\end{subfigure}
	\begin{subfigure}{.49\textwidth}
		\centering
		\includegraphics[width=\textwidth]{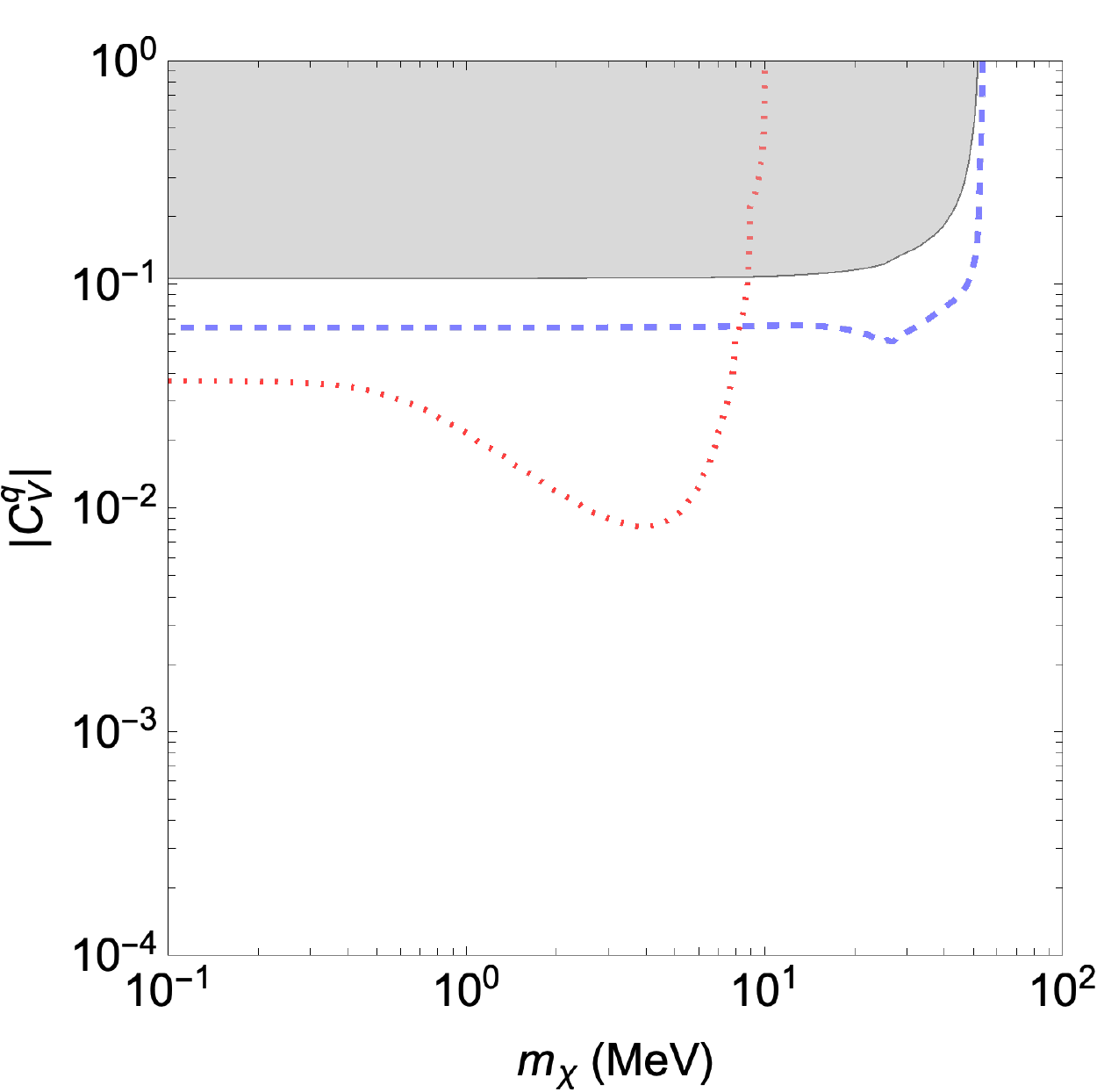}
		\label{fig:CT}
	\end{subfigure}
	\begin{subfigure}{.49\textwidth}
		\centering
		\includegraphics[width=\textwidth]{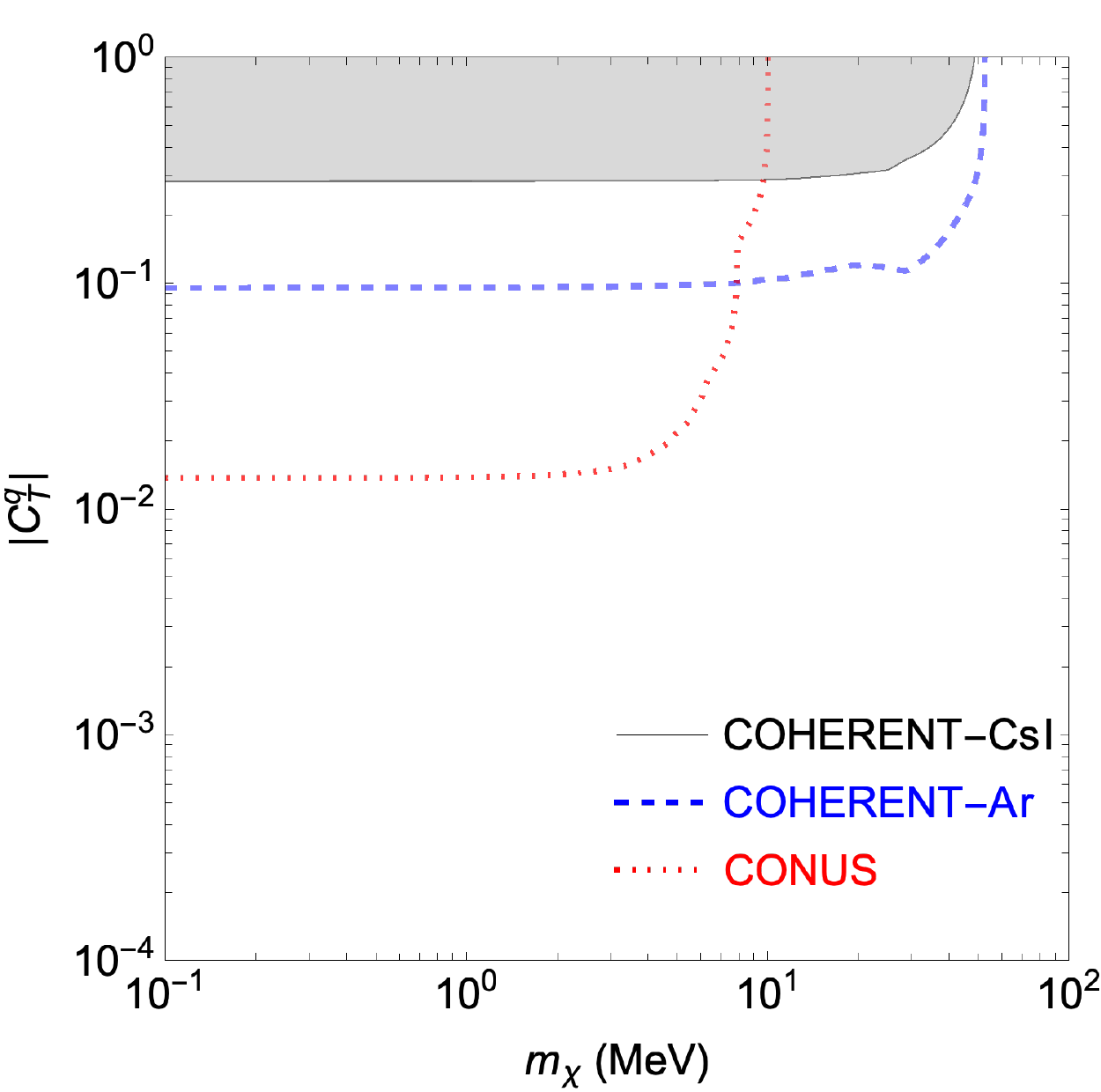}
		\label{fig:AM}
	\end{subfigure}
	\caption{The 90\% CL bounds on the Wilson coefficients as a function of the $m_\chi$ for the dipole~(upper left panel), scalar~(upper right panel), vector~(lower left panel) and tensor~(lower left panel) interactions. The gray shaded areas are obtained from the current COHERENT data with the CsI detector~\cite{Akimov:2017ade}. The blue dashed (red dotted) lines correspond to the expected limits from future COHERENT  (CONUS) experiment with an upgraded LAr (Ge) detector.}
	\label{fig:bounds}
\end{figure}

For each case, we scan over possible values of the Wilson coefficients and the $\chi$ mass $m_\chi$. Since we only consider one Wilson coefficient at a time and there is no interference term between the $\nu \Nucl\ra \chi \Nucl$ and the SM $\nu \Nucl\ra \nu \Nucl$ processes, we only place bounds on the the magnitude of the Wilson coefficients, and these bounds are not affected by their CP phases. The 90\% CL exclusion regions in the plane of $m_\chi$ versus the magnitude of Wilson coefficient in each case are shown as the gray regions in Fig.~\ref{fig:bounds} for the four cases. From Fig.~\ref{fig:bounds}, we see that the COHERENT data can only set upper bounds on the Wilson coefficients for $m_\chi\lesssim 53$~MeV. This can be understood from the kinematic constraint from Eq.~(\ref{eq:EminT}). After  marginalizing over $T$, the kinematic constraint becomes~\cite{Brdar:2018qqj},
\begin{equation}
E_\nu\geq m_\chi+\frac{m_\chi^2}{2M}\,.
\label{eq:Emin}
\end{equation}
Since the energy of neutrinos at SNS is smaller than $m_\mu/2$, we have
 $m_\chi<\sqrt{M(m_\mu+M)}-M \approx53$~MeV for the COHERENT bounds.
From Fig.~\ref{fig:bounds}, we see that the COHERENT bounds on the Wilson coefficients become flat for $m_\chi<10$~MeV.  This can be understood from Eq.~(\ref{eq:xsec}) since the terms related to $m_\chi$ are negligible when $m_\chi\ll \sqrt{MT}$.
Comparing the four cases, we see that the bound on the dipole coefficient is the most stringent and can reach as low as $3\times10^{-4}$ at 90\% CL for  $m_\chi<10$~MeV, while the bound on the tensor coefficient is the weakest and almost three orders of magnitudes weaker than the bound on the dipole coefficient.

There are several phases for future upgrades of the current detectors at the COHERENT experiment~\cite{Akimov:2018ghi}.  In order to constrain the new interactions from a future COHERENT experiment, we consider an upgrade of the LAr detector with a fiducial mass $m_{\rm{det}}=610$ kg~\cite{Akimov:2019xdj} and located at $L = 29$ m from the source. We assume 4 years of data collection with the same neutrino production rate as the current setup, which corresponds to 8.2$\times 10^{23}$ protons-on-target (POT) in total. To estimate the projected sensitivities at the LAr detector, we simulate the number of event predicted in the SM in each nuclear recoil energy bin, with the bin size being 2 keV in the range of 20 keV $ < T <100 $ keV.
For the steady-state background, we assume it is uniform in energy and the total is 1/4 of the SM expectation. We also adopt the normalization uncertainty to be 17.5\%, which includes the neutrino flux uncertainty (10\%), form factor uncertainty (5\%), signal acceptance uncertainty (5\%), and a QF uncertainty of 12.5\%~\cite{Han:2019zkz}. The projected limits are shown by the dashed lines in Fig.~\ref{fig:bounds}.  We see that future COHERENT experiment with an upgraded LAr detector can improve the current bounds by about a factor of $2-3$.
%

Since the LAr detector has a low energy threshold of 20 keV, which largely limits its ability to constrain the new interactions at low nuclear recoil energies.
Here we also explore the limits from the CONUS experiment, which utilizes a Ge detector with a very low energy threshold.\footnote{Other proposed reactor experiments such as CONNIE~\cite{Aguilar-Arevalo:2016khx} and MINER~\cite{Agnolet:2016zir} can also probe these new physics scenarios. CONNIE and MINER will utilize a smaller detector as compared to CONUS. However, depending on their experimental configurations and running time, they can reach a similar sensitivity to new physics as CONUS~\cite{Miranda:2020zji}. } The CONUS experiment measures reactor antineutrinos from a 3.9 GW nuclear power plant in Brokdorf, Germany, and the distance between the detector and reactor is 17 meters.
The current Ge detector contains only 4 kg natural Ge, and does not yield significant limits in our scenarios. Here we consider a future upgraded Ge detector with a mass of 100~kg natural Ge. The contributions of each Ge isotope are weighted by its relative abundance.
We also assume the nuclear recoil energy threshold is improved down to 0.1 keV, and take the energy bin from 0.1 keV to 2.0 keV with a bin width of 0.1 keV. We adopt the reactor flux calculated in Refs.~\cite{Huber:2011wv, Mueller:2011nm} with a conservative 5\% flux uncertainty, and assume
the background event rate to be 1 count/(day $\cdot$ keV$\cdot$ kg). After a 5 years of data collection, the CONUS bounds on the Wilson coefficients for the new interactions are shown as the red dotted curves in Fig.~\ref{fig:bounds}. We see that future CONUS experiment can set stronger bounds than future COHERENT experiment for $m_\chi $ below 10 MeV. However, the constraints from the CONUS experiment diminishes for $m_\chi>10$~MeV because the flux of reactor neutrinos drops down quickly at high energies.
Also, from the bottom left panel of Fig.~\ref{fig:bounds}, we see that the bound on the vector interaction strength becomes weak for $m_\chi \lesssim 1~\mmev$.
This can be understood from Eqs.~(\ref{eq:xsec}) and~(\ref{eq:cevns-xsec}) since the vector interaction of the process $\nu \Nucl\ra \chi \Nucl$ yields a similar dependence on $T$ as the SM process of $\nu \Nucl\ra \nu \Nucl$ when $m_\chi\ll\sqrt{2MT}$.
Moreover,  we notice that there is a kink on the bound on the dipole interaction strength at around $m_\chi\sim 4$ MeV. This is due to the partial cancellation in the differential cross section in Eq.~(\ref{eq:xsec}), and can be qualitatively explained as following.
First of all, for the CONUS experiment,  the nuclear recoil energy threshold is $0.1$~keV, $M= 7.26\times 10^5$ MeV, and $T\lesssim 3\times 10^{-3}$ MeV.
Then from Eq.(\ref{eq:EminT}), one has
\beq
E_{min}\simeq  \left[ 1.9 \left( \frac{T}{0.1\mkev} \right)^{\frac{1}{2}} + \frac{m_\chi^2}{7.5} \left( \frac{T}{0.1\mkev} \right)^{-\frac{1}{2}} \right] \mmev\,,
\eeq
for a given $T$ and $m_\chi$.
Thus, only the neutrino flux of $E \gtrsim 2~\mmev$ can contribute to the coherent scattering.
Since the typical nuclear power plant neutrino energy spectrum diminishes exponentially as the energy increases, one can take the benchmark point $\bar{E}\sim (2-3)$~MeV for a ballpark estimation\footnote{ Indeed, about $57\%$ of the effective CONUS neutrino flux ( $E>2~\mmev$) falls in the energy range of $(2-3)$~MeV, and the average neutrino energy for $E>2$ MeV is $\langle E\rangle=3.1$~MeV. }.
The leading terms of the dipole interaction differential cross section, Eq.(\ref{eq:xsec}), behaves as
$ \frac{d \sigma_{ \mbox{\tiny dip}}}{d T} \propto \left[ 1- m_\chi^2 /( 4 E^2)\right]$.
Thus, the dominant part of the event rate, see Eq.(\ref{eq:recoil-spectrum}),
\beq
{ d N_{ \mbox{\tiny dip}}\over dT } \propto  \frac{1}{T}\times\phi_\nu( \bar{E}) \times \left( 1-{m_\chi^2 \over 4  \bar{E}^2 } \right)\,,
\eeq
almost vanishes around $m_\chi\sim 2  \bar{E}\sim (4-6)\mmev$ and weakens the bound there.

\begin{figure}[t]
	\centering
	\begin{subfigure}{.49\textwidth}
		\centering
		\includegraphics[width=\textwidth]{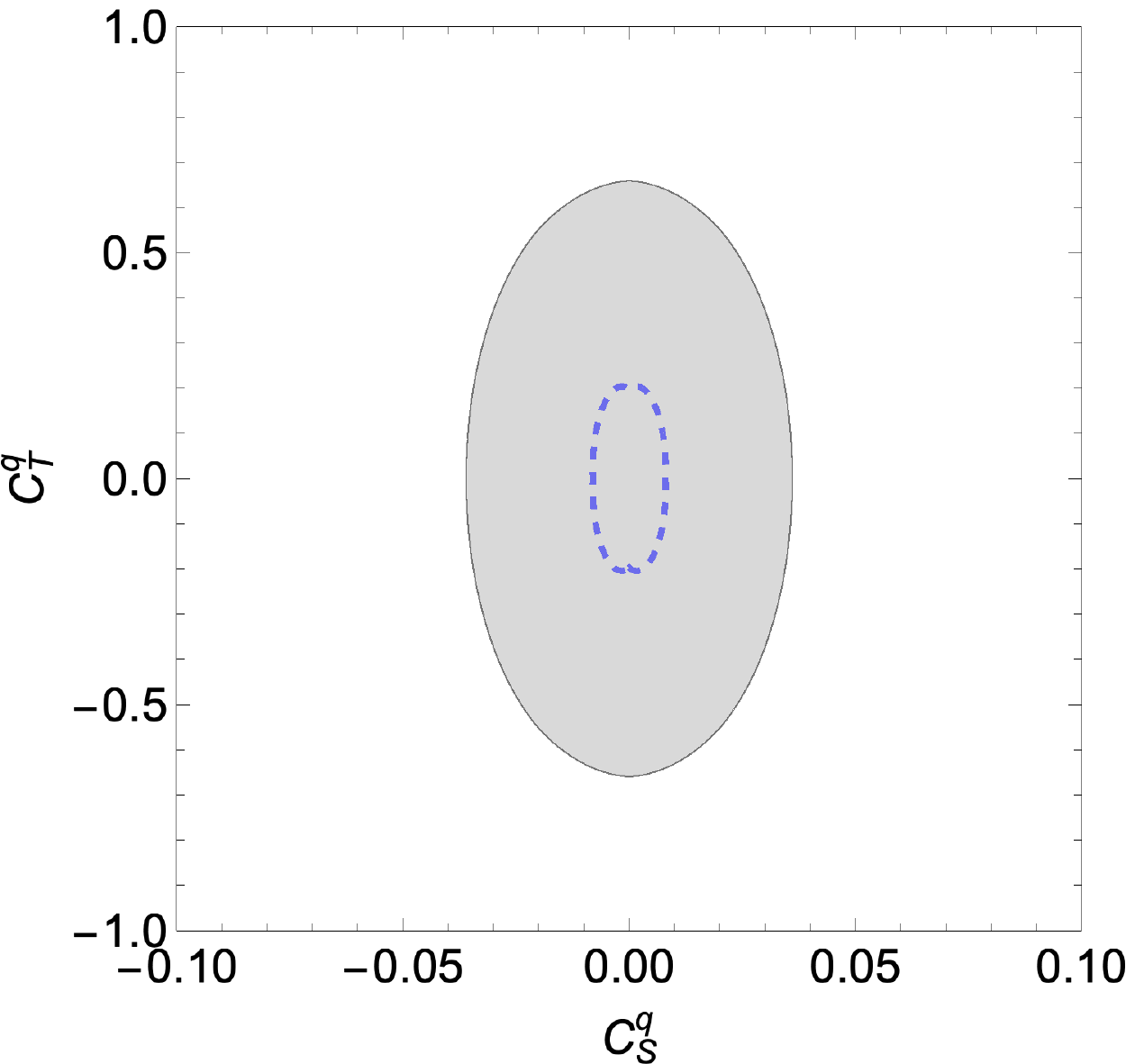}
		\label{fig:CS-CT}
	\end{subfigure}
	\begin{subfigure}{.49\textwidth}
		\centering
		\includegraphics[width=\textwidth]{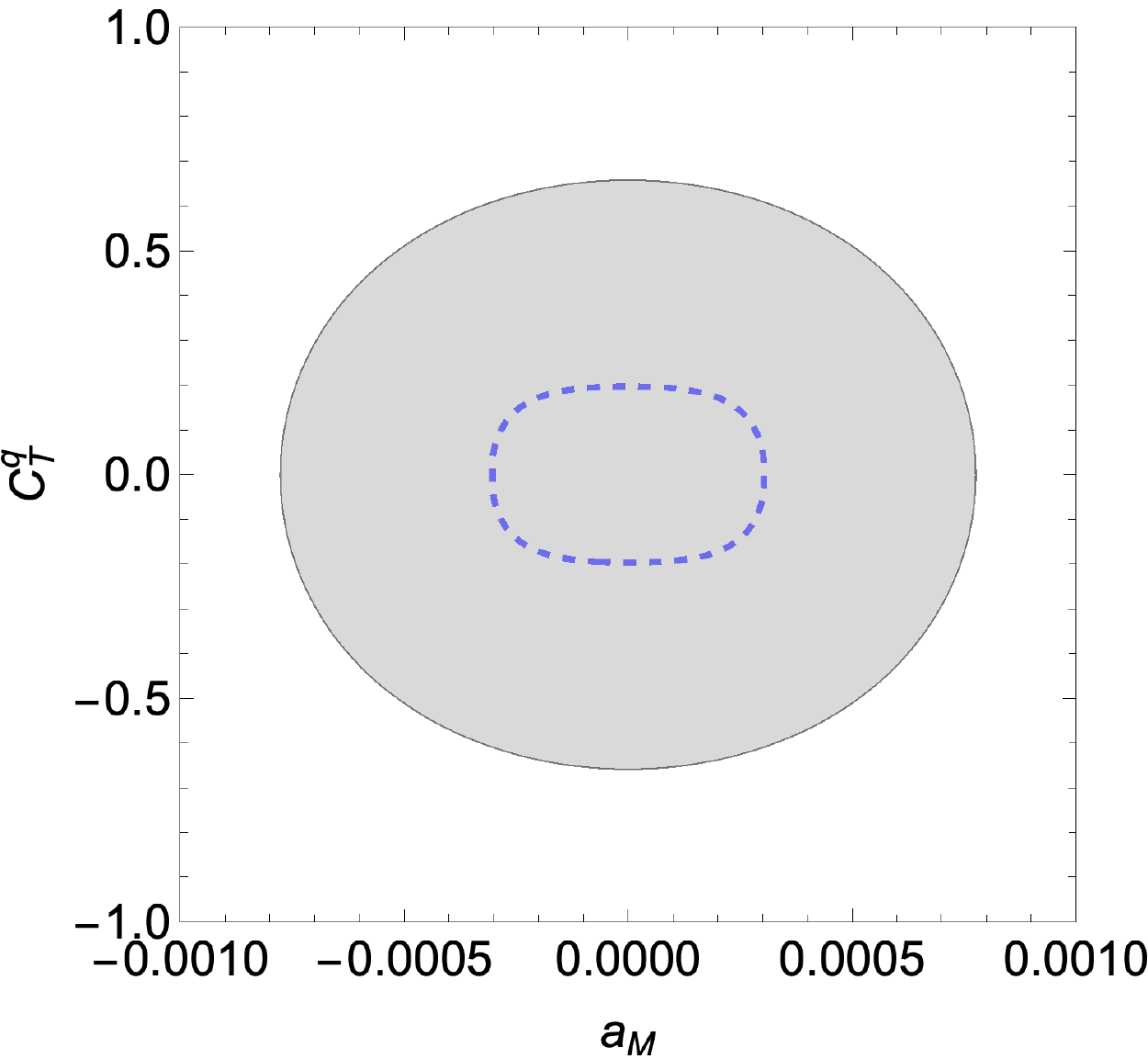}
		\label{fig:CT-AM}
	\end{subfigure}
	\begin{subfigure}{.49\textwidth}
		\centering
		\includegraphics[width=\textwidth]{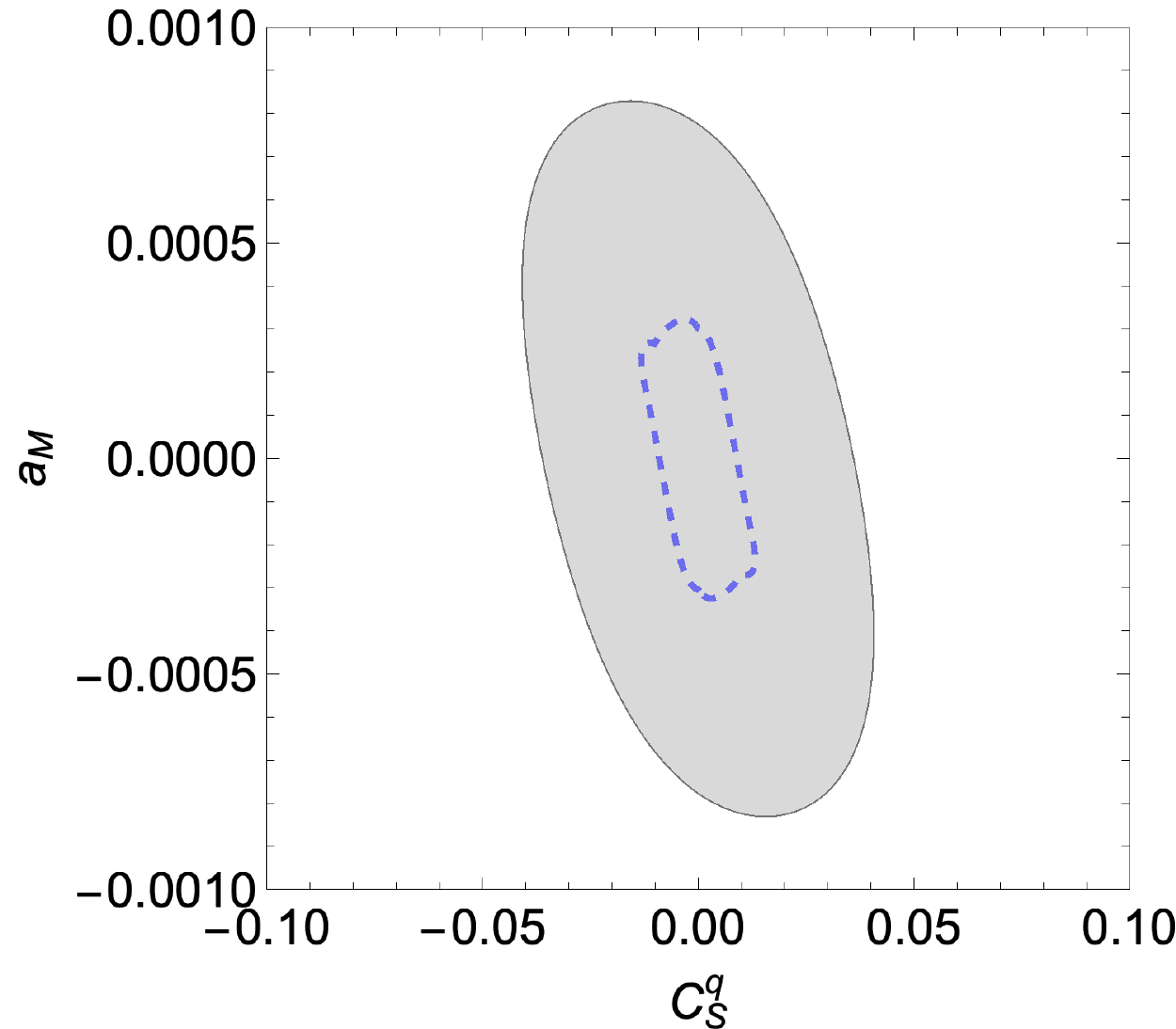}
		\label{fig:CS-AM}
	\end{subfigure}
	\caption{The $90\%$ CL allowed regions in the Wilson coefficient parameter space at the COHERENT experiment for $m_\chi=40$~MeV. The gray shaded areas are obtained from the current COHERENT CsI data, and the blue dashed curves enclose the expected allowed regions from future COHERENT experiment with an upgraded LAr detector.}
	\label{fig:correlation}
\end{figure}

Next, from Eq.~(\ref{eq:xsec}),  we see that there are interference terms among the three Wilson coefficients $C_T$, $C_S$ and $A_M$\footnote{The interference term between $C_S$ and $A_M$ is similar to that between $D_P$ and $A_E$. }. Hence, we study possible correlations between the three Wilson coefficients at the COHERENT experiment. Ideally, degeneracies between the Wilson coefficients are broken since they have different dependencies on the nuclear recoil energy in the non-interference terms. However, weak correlations may exist for some parameter values due to large systematic uncertainties.
For illustration, we choose  $m_\chi=40$~MeV, and scan the parameter space between each set of the two Wilson coefficients in $\{a_M, C^q_S, C^q_T \}$ . Note that the CONUS experiment has no sensitivity for such large $m_\chi$, and we only consider the current and future COHERENT experiment. Here we assume the Wilson coefficients are real, and allow them to be both positive and negative.  Due to the suppression of the interference terms, the allowed regions of the  Wilson coefficients will be different if nonzero CP-violating phases are taken into account. The $90\%$ CL allowed regions in the parameter space from the current CsI (future LAr) data are shown as the regions enclosed by the black solid (blue dashed) curves in Fig.~\ref{fig:correlation}. We see that the correlations between $C^q_S$ and $C^q_T$ (or $a_M$ and $C^q_T$) are very weak, but there is a slight degeneracy between $C^q_S$  and $a_M$.
The results can be understood from Eq.~(\ref{eq:xsec}). The correlation between $C^q_S$ and $C^q_T$ ($a_M$ and $C^q_T$) is negligible because the interference term is largely suppressed by  $T/E$ in the last (antepenultimate) line in Eq.~(\ref{eq:xsec}). Also, from Eq.~(\ref{eq:Emin}) and the flux given in Eq.~(\ref{eq:nu-spectra.COHERENT}), we know that for $m_\chi=40$~MeV, the $\nu  \Nucl \rightarrow \chi \Nucl$ process at COHERENT is dominated by the $\bar{\nu}_\mu$ events. Since the penultimate line in Eq.~(\ref{eq:xsec}) changes sign for antineutrinos, an anti-correlation between $C^q_S$  and $a_M$ agrees with the result shown in Fig.~\ref{fig:correlation} .

\section{Discussion and Conclusion}
\label{sec:discuss}
Before we conclude, we want to emphasize that these bounds we obtained from CE$\nu$NS are interesting for two reasons.
First of all, the process is sensitive to all singlet fermions lighter than $\sim 40$ MeV.
Secondly, as long as $m_\chi\lesssim 40$~MeV,  the constraints we obtained from COHERENT experiment  are not very sensitive to $m_\chi$.
Since there is no interference among the different singlets and the SM neutrinos, the limits we obtained for one $\chi$ can be equivalently interpreted as the limits on contributions sum of all,  say  $n_\chi$ in total,  singlet fermions lighter than $\sim 40$ MeV.

From the scan (see Fig.~\ref{fig:bounds}), barring the small correlation effects due to mutual cancellation, we have
\beqa
&& \sum_{i=1}^{n_\chi} (|a^{i}_M|^2+|a^{i}_E|^2)  \lesssim 7.2\times 10^{-8}\, (8.3\times 10^{-9})\, [1.0\times 10^{-12}]\,,\\
&& \sum_{i=1}^{n_\chi} (|C_{S i}^q|^2+|D_{P i}^q|^2)  \lesssim  4.5\times 10^{-4} (1.4\times 10^{-5})\, [4.6\times 10^{-8}]\,,\\
&& \sum_{i=1}^{n_\chi} (|C_{V i}^q|^2+|D_{A i}^q|^2) \lesssim  1.1\times 10^{-2} (4.2\times 10^{-3})\, [1.4\times 10^{-3}]\,,
\label{eq:CVbounds}\\
&& \sum_{i=1}^{n_\chi} |C^q_{T i}|^2 \lesssim  7.9\times 10^{-2} (9.1\times 10^{-3})\, [2.0\times 10^{-4}]\,,
\eeqa
at 90\% CL for $m_\chi\lesssim0.5$~MeV by using the current COHERENT (future COHERENT) [future CONUS] data.
For other mass ranges, they can be easily read from Fig.~\ref{fig:bounds}.
We stress again that the singlet fermions in our analysis need not to be the dark matter candidate. Hence, the above bounds are general and apply to any model. In particular, they are independent to those with the assumption that $\chi$ is the dark matter. For instance, the cosmic gamma-ray line background can only set a limit on the dark matter singlet dipole interaction strength, but it has no say on any unstable or short-lived singlets heavier than the DM singlet.  Also, our constraints cannot be inferred from neutrino oscillation data unless further assumption is made to relate the new physics to the SM sector\footnote{ See Appendix~\ref{apd:UVmodel} for a toy UV complete model to illustrate the physics.}.

Finally, since our constraints on the dipole interaction are much more stringent than the other interactions, one may wonder
whether it is sensitive to the dipole interaction generated by 1-loop corrections from other 4-Fermi $\nu\chi q\bar{q}$ operators.
 If we denote the dimensionless Wilson coefficient as $\tilde{C}$ as in Eq.(\ref{eq:H_eff}), a simple dimension analysis leads to
\beq
 a_{E,M}^{1-loop} \sim \tilde{C}G_F { m_q m_\chi\over 16\pi^2} \log\frac{M_W}{m_q} \simeq \tilde{C}\times 2\times 10^{-9} \times \left(\frac{m_q}{m_b}\right)\times \left(\frac{m_\chi}{\mbox{MeV}}\right)\,.
 \eeq
In the above ballpark estimation, one power of $m_\chi$ is required to flip the chirality to make the dipole interaction.
In addition,  $m_q$, the SM quark mass running in the loop, is called for to balance the dimensionality.
Therefore,the loop-generating $a_{E,M}^{1-loop}$ is too small, and the tree-level constraints on other 4-fermi $\nu\chi q\bar{q}$ operators, although weak, still matter.

In summary, we have considered the potential to probe the light singlet fermions and their effective interactions with SM quark sector by the current and planned COHERENT and CONUS experiments. The analysis is based on a model-independent dim-6 effective Lagrangian.
We find the current constraints from the COHERENT data, although loose, are profound already and complementary to the neutrino oscillation and collider measurements.  Future upgraded COHERENT and CONUS experiments will largely improve the sensitivity to new interactions, which allows us to see more details on the limits.
 We find that there is a small kink on the CONUS bound on the dipole interaction strength at $m_\chi\sim 4$~MeV which arises due to partial cancellation in the differential cross section. Also, the CONUS bound on the vector interaction strength becomes weak for $m_\chi \lesssim 1~\mmev$.
 The precise determination of the differential cross-section of coherent scattering is needed to disentangle the contribution from each effective operator, and we will leave the detailed studies to future works.

\begin{acknowledgments}
 WFC is supported by the Taiwan Ministry of Science and Technology under
Grant No. 106-2112-M-007-009-MY3. JL is supported by the National Natural Science Foundation of China under Grant No.~11905299.
\end{acknowledgments}

\appendix

\section{Tree-level amplitude}
\label{apd:amplitude}
For completeness, here we collect some calculation details of   the  tree-level $\nu(p_1) q(k_1) \ra \chi(p_2) q(k_2)$ elastic scattering cross-section.
One should keep in mind that, in reality, nucleus is the target, and the quark contribution should be summed coherently.
Moreover, nucleus mass should be used and the couplings at the quark level should be carefully replaced by the relevant form factors as discussed in Sec.\ref{sec:XS}.

The kinematics of this fundamental $2\ra 2$ process can be easily worked out as follows.
In the lab frame, the target quark of mass $M_q$ is at rest and $k_1=(M_q,0,0,0)$.
The incoming neutrino 4-momentum is denoted as $p_1=(E,E,0,0)$, and  $k_2=(M_q+T,p\cos\theta,p\sin\theta,0)$
is for the scattered quark with recoil energy $T$ and scattering angle $\theta$.  We use  $t\equiv p_1-p_2$ to denote the momentum transfer.
From the on-shell conditions $k_1^2=k_2^2=M_q^2$ and $p_2^2=m_\chi^2$, one gets $p=\sqrt{T(T+2M_q)}$
and $t^2=-2M_q T$. Also, the scattering angle can be expressed in terms of $E$ and $T$ as
\beq
\cos\theta = { T(M_q+E)+m_\chi^2/2\over E\sqrt{T(T+2M_q)}}\,.
\label{eq:angle}
\eeq
For a given $T$, the minimal energy required to generate the elastic scattering is thus
\beq
E_{min}={ T M_q + m_\chi^2/2\over \sqrt{T(T+2M_q)}-T}\,.
\label{eq:EminT}
\eeq
And it can be proved that for the physical $T$, it has an extreme $E_{min}\geq m_\chi+m_\chi^2/2M_q$.
The other scalar products can be easily derived to be:
\beqa
p_1\cdot k_1 =M_q E\,,\,\,\, p_1\cdot k_2=M_q(E-T)-m_\chi^2/2\,,\,\,\, p_1\cdot p_2 =M_qT+m_\chi^2/2\,,\nonr\\
p_2\cdot k_1 =M_q (E-T)\,,\,\,\, p_2\cdot k_2=M_q E-m_\chi^2/2\,,\,\,\, k_1\cdot k_2 =M_q(M_q+T)\,.
\eeqa

From the effective Lagrangian, Eq.(\ref{eq:H_eff}), the tree-level Feynman diagrams for the process $\nu(p_1) q(k_1) \ra \chi(p_2) q(k_2)$ are shown in Fig.\ref{fig:TLFD}.
The amplitude is given by
\beqa
i{\cal M} &=& \frac{i G_F}{\sqrt{2}} \left[\bar{u}(p_2) \gamma^\mu(C_V^* +\gamma_5 D_A^*) u(p_1)\right]\left[\bar{u}(k_2) \gamma_\mu u(k_1)\right]\nonr\\
&+&\frac{i G_F}{\sqrt{2}} \left[\bar{u}(p_2) (C_S^* +i \gamma_5 D_p^*) u(p_1)\right]\left[\bar{u}(k_2)  u(k_1)\right]\nonr\\
&+&\frac{i G_F}{\sqrt{2}} C_T^* \left[\bar{u}(p_2) \sigma^{\mu\nu} u(p_1)\right]\left[\bar{u}(k_2) \sigma_{\mu\nu} u(k_1)\right]\nonr\\
&-&\frac{ G_F v_H Q_q |e| }{t^2} \left[\bar{u}(p_2) \sigma^{\mu\nu} (a_M^* +i \gamma_5 a_E^*) u(p_1)\right]\left[\bar{u}(k_2) \gamma_\mu t_\nu u(k_1)\right]\,,
\eeqa
where $Q_q|e|$ is the electric charge of quark $q$, and all the couplings are at the quark level and flavor-dependent.
The calculation of the amplitude squared is straightforward.
Then, from the average amplitude squared, one obtains the differential cross section
\beq
\frac{d \sigma}{ d T} ={ \left\langle |{\cal M}|^2 \right\rangle\over 32\pi M_q E^2}\,.
\eeq

\section{A UV complete model}
\label{apd:UVmodel}
To further illustrate the physics discussed above, let us consider the coherent scattering implication to a UV complete model.
Our custom-made toy model  is a simple extension of the type-I seesaw model with total $n$ right-handed sterile neutrinos.
The model Lagrangian is trivial and will not be spelled out here.
We denote the mass eigenstates as $\tilde{ \nu}\equiv \{ \nu_1,\nu_2,\nu_3, \nu_4,\cdots,\nu_{3+n}\}$, where $\nu_{1,2,3}$ are the sub-eV light active neutrinos.  For the flavor basis,  the notation $\tilde{N}\equiv \{ \nu_e,\nu_\mu,\nu_\tau, \chi_1,\cdots,\chi_{n}\}$ is adopted.
We assume that three out of the $n$ sterile neutrinos, $\nu_{1+n,2+n,3+n}$, are the heavy ones, decoupled at the low energies, as in the classic high scale type-I seesaw. Moreover, with some parameter turning, the details are not important here, all the other sterile neutrinos, $\nu_{4,\cdots,n}$, acquire their masses, $m_{4,\cdots,n}$,  in the range of $(1-40)$ MeV\footnote{ Here, dark matter is not our concern.}.
The mass and flavor states are related by an unitary transformation, $ \tilde{N}=U \tilde{\nu}$, which diagonalizes the neutral fermion mass matrix.

The $\nu_\mu\ra \nu_e$  transition probability in vacuum is given by
\beq
P_{\nu_\mu\ra \nu_e}(L) =\left| \sum_{i=1}^n U_{\mu i}\, \exp \left(\frac{-i (\triangle m_i)^2 L}{2E_\nu}\right)\, (U^\dag )_{ie}\right|^2\,,
\eeq
where $E_\nu$ is the neutrino source energy, $L$ is the distance neutrinos travel from the source, and $\triangle m_i^2 \equiv m_i^2 -m_1^2$.
At the near detector, $L \simeq 0$, the matter effect is negligible and the exponential factors can be dropped. By unitarity, the probability becomes
\beq
P_{\nu_\mu\ra \nu_e}(L=0) \simeq \left| -\sum_{ i=n+1}^{n+3} U_{\mu i}(U^\dag )_{ie}\right|^2\,.
\label{eq:NUniBound}
\eeq
From neutrino oscillation data only, the current upper bound on this quantity  is $1.1\times 10^{-3}$ at 90\%~CL\cite{Avvakumov:2002jj}, and could be pinned down to the ${\cal O}(10^{-5})$ level with the planned near detector at Fermilab\cite{Miranda:2018yym}.
Note that all information about the mixings with light sterile neutrinos does not present.

A similar bound can be inferred from the  SM $Z^0$ boson invisible decay width measurement. In this toy model, the SM $Z^0$ boson can decay into any light $\overline{\tilde{\nu}_i}\tilde{\nu}_j $ pair except the heavy three. Thus,
\beq
{\triangle \Gamma(Z^0\ra \mbox{invisible})\over \Gamma_{SM}(Z^0\ra \mbox{invisible})}\simeq
-2 \sum_{i=n+1}^{n+3}\sum_{l=e,\mu,\tau} |U_{l i}|^2\,.
\eeq
The current value of $N_\nu =2.984\pm0.008$ from LEP\cite{PDG} can be translated to
\beq
\sum_{l=e,\mu,\tau}\sum_{i=n+1}^{n+3} |U_{l i}|^2 < 1.1 \times 10^{-2}
\label{eq:ZinvBound}
\eeq
at two sigma level. Again, it is not sensitive to the properties of the light sterile neutrinos.

On the other hand, due to the mixing, the new $\nu\mhyphen\chi\mhyphen Z^0$ NC interaction exists, and the SM $Z^0$ boson is the mediator. Due to the coherent enhancement, we have roughly  $|C_{V,l i}|=|D_{A,l i}|\simeq 0.5\times (4 Z s_W^2+N-Z)|U_{l i}|\, (l=e,\mu,\tau)$ at the nucleus level. All other couplings are either generated at the loop-level or suppressed by the active neutrino masses.
Then, using Eq.(\ref{eq:CVbounds}), we have
\beq
\sum_{i=4}^n |U_{l i}|^2 \lesssim 0.65 \,(0.28) \,[0.09]
\label{eq:nNCSBound}
\eeq
at 90\% CL from the current COHERENT (future COHERENT) [future CONUS] data for this toy model.
Due to the accidentally small proton weak charge, $Q_w^p\propto 1-4s_W^2$, the above bound is much weaker than Eq.(\ref{eq:CVbounds}).
Note Eq.(\ref{eq:NUniBound}), Eq.(\ref{eq:ZinvBound}), and Eq.(\ref{eq:nNCSBound}) are constraining three different quantities and independent to each other.
And it is clear from this example that the neutrino oscillation, collider search, and CE$\nu$NS are complementary to each other.
For the illustration purpose, we further consider a universal mixing between active and sterile neutrinos, namely $|U_{l i}|$ is a constant for $i \geqslant 4$.
Then if any nonzero $P_{\nu_\mu\ra \nu_e}$ or $\triangle N_\nu$ is measured in the future, the upper limit on the number of light
sterile neutrinos, $ m_\chi \in [1, 40]$~MeV, in this toy model can be deduced from Eq.~(\ref{eq:nNCSBound}).

\end{document}